\documentclass[aps,prl,twocolumn,showpacs,superscriptaddress,groupedaddress]{revtex4}  % for review and submission
\usepackage{graphicx}  % needed for figures
\usepackage{dcolumn}   % needed for some tables
\usepackage{bm}        % for math
\usepackage{amssymb}   % for math
\usepackage{xcolor}
\usepackage{siunitx}
\usepackage{amsmath}
\usepackage{amssymb}
\usepackage{soul}

\newcommand{\xleft}{x_-}
\newcommand{\xright}{x_+}
\newcommand{\bendability}{\nu}

\begin{document}

\title{Wettability-independent droplet transport by \emph{Bendotaxis}}
\author{Alexander T. Bradley, Finn Box, Ian J. Hewitt and Dominic Vella}
\affiliation{Mathematical Institute, University of Oxford, Woodstock Rd, Oxford OX2 6GG, United Kingdom}
%\input author_list.tex       % D0 authors (remove the first 3 lines
                             % of this file prior to submission, they
                             % contain a time stamp for the authorlist)
                             % (includes institutions and visitors)
\date{\today}

\begin{abstract}
We demonstrate \textit{bendotaxis}, a novel  mechanism for droplet self-transport at small scales. A combination of bending and capillarity in a thin channel causes a pressure gradient that, in turn, results in the spontaneous movement of a liquid droplet. Surprisingly, the direction of this motion is always the same, regardless of the wettability of the channel. We use a combination of experiments at a macroscopic scale and a simple mathematical model to study this motion, focussing in particular on the time scale associated with the motion. We suggest that \emph{bendotaxis} may be a useful means of transporting droplets in technological applications, for example in developing self-cleaning surfaces, and discuss the implications of our results for such applications.
\end{abstract}

\maketitle
%\graphicspath{{./Paper_figures/}}
%\section{\label{sec:level1}First-level heading}
% sections are not used for PRL papers

Control and transport of liquid droplets on small scales, where surface forces dominate, is of critical importance in  applications including microfluidics, microfabrication and coatings~\cite{Style13PNAS, Squires05RevModPhys, Srinivasarao01Science, deGennes04, Mastrangelo93JMicroelec}. Active processes including gradients in temperature~\cite{Sammarco1999RKC,Pratap2008Langmuir}, applied electric potentials~\cite{Mugele2005CondMat} and mechanical actuation~\cite{Prakash2008Science}
have been used successfully to generate such fine scale control. Recently, however, there has been growing interest in generating droplet motion  passively. This can be achieved using a fixed geometry, in which droplets move in response to tapering~\cite{Kim11AngewandteChemie, Reyssat14JFM, Lv14PRL, Renvoise2009EPL,Guan17SoftMatter}. When the geometry is responsive (e.g.~with deformable boundaries), however, more possibilities open up, including \emph{durotaxis}~\cite{Style13PNAS} and \emph{tensotaxis}~\cite{Bueno17ExtrMech}, which rely on gradients in  stiffness and strain of an underlying soft substrate, respectively, to control motion. Here we introduce a novel, passive droplet transport mechanism that takes advantage of the capillary-induced bending of a narrow  channel whose walls are slender and hence deformable; we term this motion \emph{bendotaxis}. Importantly, we shall demonstrate that  the direction of bendotaxis  is independent of wettability. This is in contrast to durotaxis, in which wetting and non-wetting droplets have been reported to move in opposite directions~\cite{Bueno18SoftMatter}.

\begin{figure}
\centering
\includegraphics[width=0.95\columnwidth]{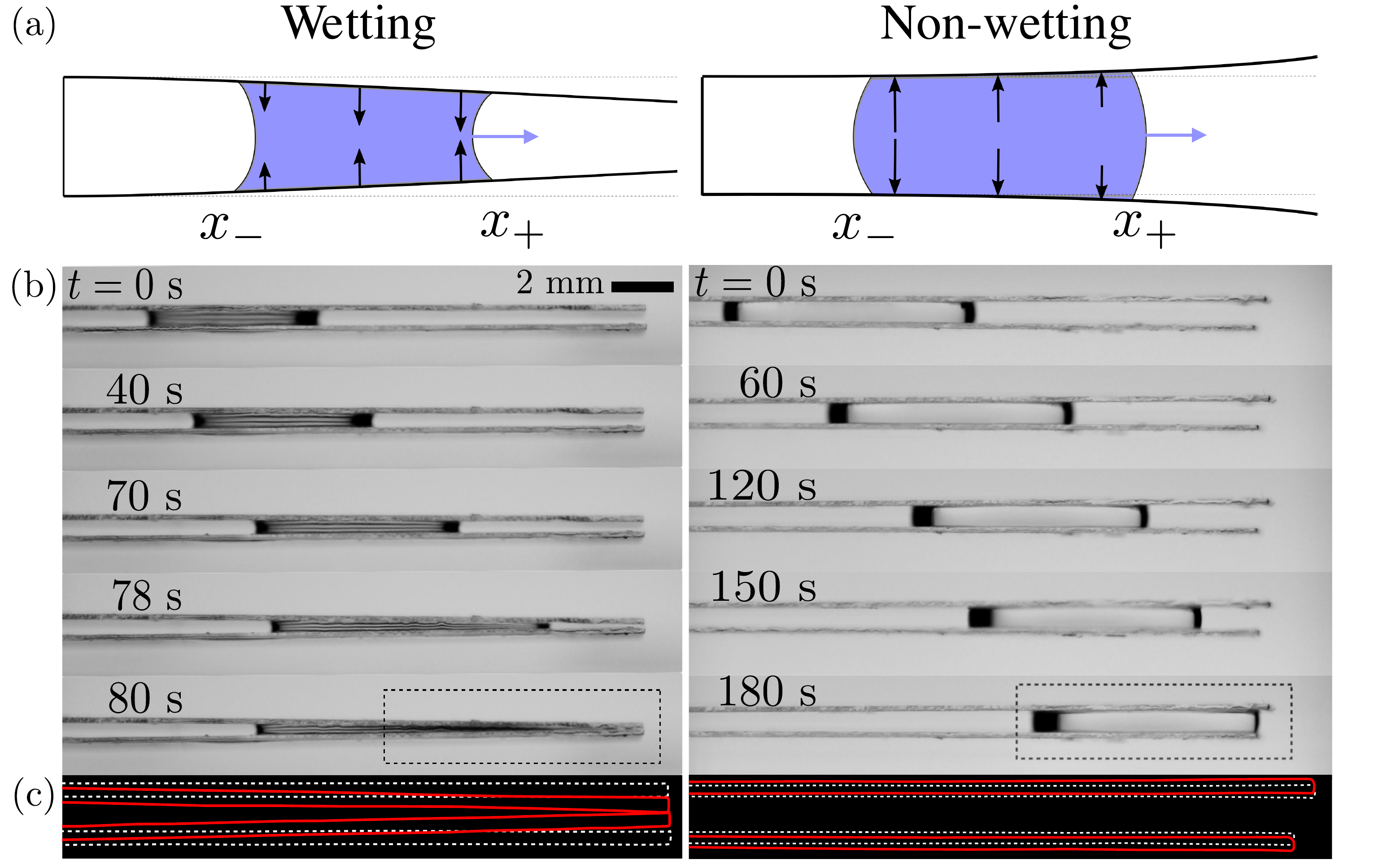}
\caption{\label{fig:fig1_ProofOfConcept}  (a) Schematic diagrams explain the mechanism behind bendotaxis for wetting (left) and non-wetting (right) drops: black arrows indicate the sign and magnitude  of the Laplace pressure within the drop; purple arrows show the direction of decreasing pressure  and hence motion. (b) Experimental demonstration of bendotaxis for a wetting silicone oil droplet (left) and a non-wetting water droplet (right), each between initially parallel, yet deformable, SLIPS. While the deformation of the channel is different in each case, the direction of droplet motion is the same.  (c) Comparison of final channel shape (red lines) with the initial channel shape (dotted white lines)
 for the section highlighted by the dashed box in (b). }
\end{figure}

Figure \ref{fig:fig1_ProofOfConcept}(a) illustrates the mechanism behind bendotaxis: two, initially parallel, bendable walls are clamped at one end and free at the other, forming a two-dimensional channel. If a wetting droplet is introduced between the walls, the negative Laplace pressure  deflects the walls inwards. The deformation is larger at the meniscus closer to the free end (referred to as $\xright$) than at the clamped end ($\xleft$). The pressure is therefore more negative at $\xright$ than at $\xleft$; the resulting pressure gradient drives the droplet towards the free end. Provided the contact angles remain the same and the beams do not touch, this motion will continue until the droplet reaches the free end. %Here, we are only concerned with configurations in which the end do not touch, discussion of the effect when beams interact can be found in (supp info). 
For a non-wetting droplet introduced into the channel, the Laplace pressure is positive, pushing the beams away from one another but the resulting pressure gradient is again negative, driving the droplet towards the free end.

This mechanism is reproducible in a simple laboratory experiment. We fabricated channels using a rigid separator and glass coverslips. %(thickness 180 $\mu$m, width 5 mm) cut from borosilicate coverslips (Agar Scientific)
Figure \ref{fig:fig1_ProofOfConcept}(b) shows time-series of a wetting silicone oil droplet  and of a non-wetting water droplet in such a channel. In both cases, the droplets move towards the free end of the channel. To observe the deflection of the coverslips, we compare their shapes  in the final configuration with those prior to the introduction of the droplet (fig.~\ref{fig:fig1_ProofOfConcept}(c)).  In the wetting case, both coverslips are deflected inwards, while in the non-wetting case both are deflected outwards, in accord with our physical description. The observed deflections also provide evidence that motion is not simply caused by the weight of the droplet, which would cause the lower coverslip to deflect downwards in both cases. (Our neglect of gravity is justified in~\cite{endnote47}.)

\begin{figure}[h!]
\centering
\includegraphics[width=0.9\columnwidth]{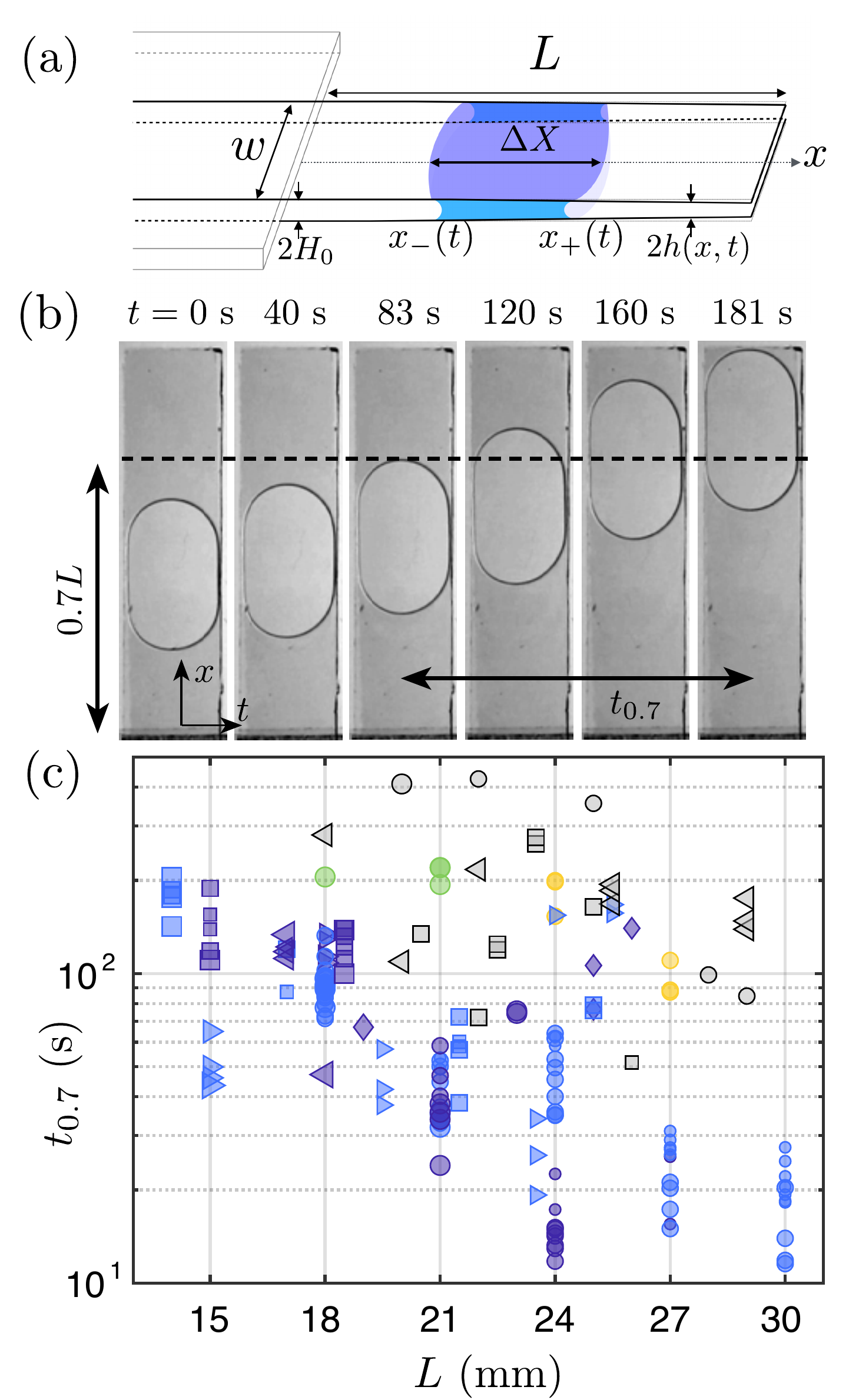}
\caption{\label{fig:fig2_experiment} (a) Schematic of a droplet  in a flexible channel of undeformed wall separation $2H_0$, width $w$ and length $L$, with  menisci positioned at $x=\xleft(t)$ and $x=\xright(t)$. (b)  Top view of a  droplet of V50 silicone oil (channel length $L = 18 \mathrm{~mm}$, width $w=5\mathrm{~mm}$, wall separation $2H_0=310\mathrm{\mu m}$, thickness $b=300\mathrm{~\mu m}$ and droplet volume $V =10~\mu$L). Although the droplet spans the width of the channel, it is not precisely two-dimensional. We  present data for $t_{0.7}$: the time between the events $\xright/L = 0.7$ and $\xright/L = 1$, which corresponds to the time between the fourth and sixth images here.  (c) Raw experimental measurements of $t_{0.7}$, for different beam lengths, $L$. Data are shown for droplets of wetting silicone oil and a non-wetting water--glycerol mix; droplet type is encoded as in fig.~\ref{fig:fig3_collapse}. ~Different shapes  encode channel wall separation as follows: $2H_0 = 310~\mu \mathrm{m} $ ($\triangleright$), $360~\mu \mathrm{m} $ ($\triangleleft$), $430~\mu \mathrm{m} $ ($\square$), $ 540~\mu \mathrm{m} $ ($\bigcirc$), $630~\mu \mathrm{m} $ ($\diamondsuit$).  The size of each point encodes the approximate fraction of the channel taken by the droplet ($\hat{V}= \Delta X/L$),  with bins corresponding  to $\hat{V} < 0.25$, $0.25 \leq  \hat{V} < 0.35$, $0.35 \leq  \hat{V} < 0.45$ and $\hat{V} \geq 0.45$.}

\end{figure}

To gain insight into the dynamics of bendotaxis we performed a series of more detailed experiments: sections of borosilicate glass coverslips (Agar Scientific, Young's modulus $E = 63 \mathrm{~GPa}$, thickness $160\mathrm{~\mu m}\leq b \leq 310\mathrm{~\mu m}~(\pm5\mathrm{~\mu m})$, width $w =  5\pm 0.5\mathrm{~mm}$) were first treated (see below) before being clamped with a  separation $310\mathrm{~\mu m}\leq 2H_0 \leq 630 \mathrm{ ~\mu m} ~(\pm5\mathrm{ ~\mu m})$ to create an open-ended channel, as in fig.~\ref{fig:fig2_experiment}(a). The channel length $14\mathrm{~mm}\leq L\leq 30\mathrm{~mm}~(\pm0.25\mathrm{~mm})$ is controlled by changing the clamping position (while maintaining a relatively long clamped section to ensure there is no intrinsic tapering, which would alter the dynamics~\cite{Renvoise2009EPL,Reyssat14JFM,Gorce16Langmuir}). 

The treatment of the glass and the droplets used depended on the required wetting conditions: for the non-wetting case, the walls were sprayed with a commercial hydrophobic spray (Soft-99, Japan) and dip-coated with silicone oil V5 (Sigma-Aldrich, USA) forming a  slippery lubricant-infused porous surface (SLIPS)~\cite{Wong2011,Smith2013,Luo17PRApplied,endnote47}. Droplets were formed from a water--glycerol mix (70$\%$ water by weight, dynamic viscosity $\mu = 36\pm5\mathrm{~mPa \ s}$, surface tension $\gamma = 67 \mathrm{~mN \, m}^{-1}$); this combination of drop and lubricant liquids ensures a large advancing contact angle ($\theta_{a} = 102\pm 2 \si{\degree}$~\cite{endnote47}), low hysteresis (receding angle $\theta_{r} = 100\pm  2\si{\degree}$) and a large enough drop:lubricant viscosity ratio  that viscous dissipation occurs primarily within the droplet~\cite{Keiser17SoftMatter,endnote47}.

 In the wetting case, we pre-wetted the glass with silicone oil  (pre-wetting was performed on both bare glass, as well as with glass pre-treated by hydrophobic spray to better retain the wetting film; we find no difference between these two cases in our experimental data \cite{endnote47}). Droplets of silicone oils V50, V100, V350 and V500, were used to vary the dynamic viscosity in the range $48 \leq \mu \leq 480 \mathrm{~mPa \ s}\pm5\%$ while maintaining  $\gamma = 22 \mathrm{~mN \, m}^{-1}$. These droplets perfectly wet the pre-wetted glass but form a  capillary bridge with well-defined menisci.

The droplet volume was systematically varied in the range $10\pm0.5\mathrm{~\mu L}\leq V \leq 25\pm0.5\mathrm{~\mu L}$, leading to different initial droplet lengths $\Delta X=\xright(0)-\xleft(0)$ and, hence, different relative volumes $\hat{V} = \Delta X/L$ (constant in each experiment). The wall separation  at the free end is enforced to be  $2H_0$ during droplet deposition but removed shortly after, which corresponds to $t = 0$; in this brief period immediately following deposition, droplet motion is negligible \cite{Mastrangelo93JMicroelec, endnote47}. The experiment is photographed from above, as  in fig.~\ref{fig:fig2_experiment}(b), and the position of the leading meniscus, $\xright(t)$, is recorded and tracked using image analysis software in \textsc{Matlab}. (Note that the droplet volumes $V$ were chosen so that the drop spans the width $w$ of the channel, becoming effectively two-dimensional, fig.~\ref{fig:fig2_experiment}(b).)

To quantify the time scale of motion in a reproducible manner (independent of the initial droplet position), we measure the time, $t_{X}$, taken for the droplet to pass from $\xright/L = X$ to the free end $\xright/L = 1$.  This quantity is approximately independent of the initial condition, provided inertia is negligible. Here we present results for $X = 0.7$, which is arbitrary but covers a significant portion of the motion for which beam bending occurs over a length comparable to $L$ (a fact used in the following scaling arguments), whilst still allowing most experiments to be included.

\begin{figure}[t]
\centering
\includegraphics[width=0.95\columnwidth]{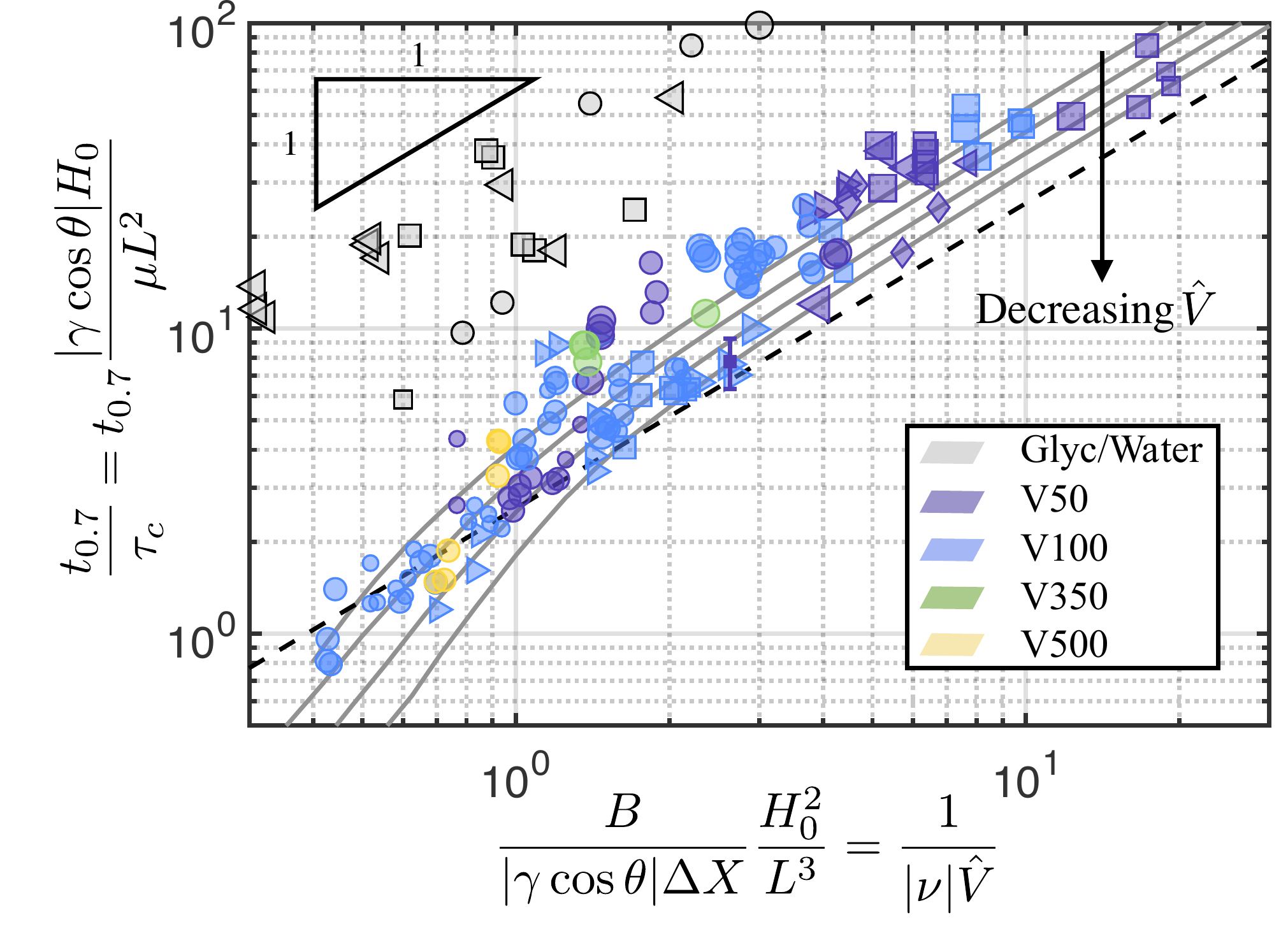}
\caption{\label{fig:fig3_collapse} Collapse of experimental data when rescaled according to (\ref{E:scaling}).  Points correspond to experimental observations (with volume and  wall separation encoded by point size and shape, respectively, as in fig.~\ref{fig:fig2_experiment}; droplet type is  encoded by color, as  indicated in the legend). The single set of error bars extends one standard deviation away from a particular data point, computed from 20 measurements, and is similar for each experiment~\cite{endnote47}. Solid curves show results from numerical solutions of \eqref{E:NumSolEq}--\eqref{E:NumSolKBC} with $\hat{V} = 0.5,0.4,0.3$ and $\hat{V} = 0.2$.  Also plotted is the asymptotic result \eqref{E:small_V_scaling} (black dashed line), valid for $\hat{V} \ll 1$ (corresponding to the upper right corner of this plot).}
\end{figure}

Raw measurements of $t_{0.7}$  are presented in fig.~\ref{fig:fig2_experiment}(c), and indicate a strong dependence on both channel geometry and droplet viscosity. To gain theoretical insight, we first consider a  scaling argument assuming a small relative volume, $\hat{V}\ll 1$, which captures the combination of elasticity and capillarity involved. Droplet motion is driven by the Laplace pressure change that results from droplet-induced tapering of the channel (we assume a constant surface tension $\gamma$ and contact angle $\theta$ at the leading and rear menisci and  neglect the surface tension from the sides, shown to be relatively unimportant in a similar situation \cite{Kwon08JApplPhys}). In a narrow channel, the pressure change across the droplet due to a tapering angle $\alpha$ can be approximated as $
\Delta P \sim  \alpha \gamma \cos \theta \Delta X / H_0^2$. Since the channel walls bend over a length comparable to $L$ (provided the drop is relatively far from the clamp), but are only subject to a Laplace pressure over the length of the drop, linear beam theory~\cite{Howell08} suggests that $\alpha \sim \gamma \cos \theta ~L^2 \Delta X/B H_0$.
(Here $B = Eb^3/12$ is the bending stiffness of the wall per unit width \cite{audoly10,endnote47}.) Therefore, the pressure gradient over the (small) droplet is estimated as
\begin{equation}
\frac{\partial P}{\partial x} \sim \frac{L^2}{H_0^3}\frac{\gamma^2 \cos^2 \theta \Delta X}{B}.
\end{equation}
Lubrication theory~\cite{Leal07} provides the timescale for a droplet of viscosity $\mu$ to move along the length of the beams as $\tau \sim \mu L / (H_0^2  P_{x})$. When considered relative to a capillary timescale $\tau_{c} = \mu L^2/(\left|\gamma\cos\theta\right| H_0) $, this yields
\begin{equation}\label{E:scaling}
\frac{\tau}{\tau_c}=\frac{\tau \left|\gamma \cos \theta \right| }{\mu } \frac{H_0}{L^2} \sim \frac{B}{\left|\gamma \cos \theta \right| \Delta X }  \frac{H_0^2}{L^3}.
\end{equation}
The scaling \eqref{E:scaling}  provides a reasonable collapse of the experimental data for all of the wetting data, see fig.~\ref{fig:fig3_collapse} and fig.~S1 \cite{endnote47}. However, the non-wetting experiments show two families with  a similar scaling trend but modified prefactors, which may be due to a change in the effective value of $\left| \gamma \cos \theta\right|$ between measurements made on a single SLIPS and  experiments in a narrow channel. A possible cause for such a change in the driving Laplace pressure is a thin oil layer `cloaking'  the droplet \cite{Schellenberger15SoftMatter}. Quantifying this is beyond the scope of the present study, but we note that the discrepancy in prefactor would be eliminated by a relatively small change in the effective contact angle of $\lesssim7^\circ$.

For moderate to large values of the abscissa in fig.~\ref{fig:fig3_collapse} we observe the linear scaling of \eqref{E:scaling} (valid for $\hat{V}\ll1$). However, at smaller values (larger $\hat{V}$) the linear scaling appears to break down. To go beyond this scaling argument and determine the effect of finite droplet volumes $\hat{V}$, we formulate a detailed mathematical model. Combining lubrication theory and linear beam theory (and neglecting the weight and tension within the beam) leads to a nonlinear PDE for the deformed shape of the beams $h(x,t)$ within the wetted region~\cite{Aristoff11JNonlinMech, Taroni12JFM, Duprat11JFM}:
\begin{equation}\label{E:NumSolEq}
\frac{\partial h}{\partial t} = \frac{B}{3 \mu}\frac{\partial}{\partial x}\left[ h^3 \frac{\partial^5 h}{\partial x^5}\right]\;, \qquad \xleft(t) < x < \xright(t).
\end{equation}
The shape of the channel wall  out of contact with the droplet satisfies $\partial^4h/\partial x^4=0$ and depend on time only through the meniscus positions. At each meniscus we require continuity of shape, slope, moments and shear force, consistent with the assumption of a small aspect ratio, $H_0/L \ll 1$, used in lubrication theory~\cite{Taroni12JFM}. The pressure jump between dry and wet portions of the beam is due to the Laplace pressure at the meniscus, so that
\begin{eqnarray}
B\left.\frac{\partial^4h}{\partial x^4}\right|_{x=x_m} = -\frac{\gamma \cos \theta}{h(x_m,t)}, \qquad x_m=\xleft\;,  \ \xright\;.\label{E:NumSolBC4}
\end{eqnarray}
As before, we have assumed that the contact angles at the advancing and receding menisci are equal and constant. On the timescales considered here, evaporation is negligible \cite{endnote47} conservation of mass then requires
\begin{eqnarray}\label{E:NumSolKBC}
\frac{\mathrm{d}x_m}{\mathrm{d}t}  = - \frac{B h^2}{3\mu}\left.\frac{\partial^5 h}{\partial x^5}\right|_{x = x_m}, \qquad x_m = \xleft, \xright,
\end{eqnarray} 
We apply clamped boundary conditions at $x=0$, while the end $x=L$ is free, i.e.~$h(0,t)=H_0$, $h_x(0,t)=0$, and $h_{xx}(L,t)=h_{xxx}(L,t)=0$. The problem is closed by specifying initial conditions for the beam shape, $h(x,0)=H_0$, and the meniscus positions $\xleft(0) = \xleft^0$, $\xright(0) = \xright^0$.

Asymptotic analysis of the problem (\ref{E:NumSolEq})-(\ref{E:NumSolKBC}) for $\hat{V} \ll 1$ shows that the beam deflection is small and that the drop length is approximately constant throughout the motion~\footnote{See Supplementary Information, which includes refs \cite{SchottGlassData,Takamura2012,Schneider12NMeth,Yuan2013contact,Onda1996Langmuir,Canny86,Hu2002JPhysChemB,Zhornitskaya00,Shampine07TOMS}, for details of the  experimental procedure, asymptotic analysis and numerical techniques.}. The evolution of the meniscus positions are then governed by the ODEs
\begin{equation}
\frac{\mathrm{d} x_{m}}{\mathrm{d}t} = \frac{ \gamma^2 \cos^2 \theta \Delta X}{6 B\mu  H_0}x_{m}^2\;, \qquad x_{m} = \xleft, \xright.
\end{equation} The ODE for $\xright(t)$ may be solved to  give the time $t_X$ taken to move from $\xright/L=X$ to $\xright/L=1$ as
\begin{equation}\label{E:small_V_scaling}
\frac{t_{X} }{\tau_c} = \frac{6(1-X)}{X}\frac{B}{\left|\gamma \cos \theta\right| \Delta X}\frac{H_0^2}{L^3}.
\end{equation} Eqn \eqref{E:small_V_scaling} confirms the scaling result \eqref{E:scaling} and provides the appropriate pre-factor, which, with $X=0.7$, corresponds to the black dashed line  in fig.~\ref{fig:fig3_collapse}.

To facilitate numerical solutions of the full problem (\ref{E:NumSolEq})-(\ref{E:NumSolKBC}), we non-dimensionalize axial lengths by $L$, the beam deformation by $H_0$ and time by the capillary timescale  $\tau_{c} $, introduced earlier. In addition to the relative volume $\hat{V}$, we identify a further dimensionless parameter
\begin{equation}
\bendability = \frac{L^4 \gamma \cos \theta}{H_0^2 B}\;,
\end{equation} which represents the ability of the droplet surface tension to bend the channel walls. We refer to the parameter $\bendability$ as a channel `bendability', though it is also related to the reciprocal of the elastocapillary number  \cite{Mastrangelo93JMicroelec}. Note that wetting drops have $\bendability > 0$ while non-wetting drops have $\bendability < 0$, consistent with the sign of the pressure in equation \eqref{E:NumSolBC4}.

The problem is fully specified by the values of $\bendability$, $\hat{V}$, and the initial condition $\xright^0/L$, and can be solved numerically in \textsc{Matlab} using the method of lines~\cite{Schiesser91,endnote47}. The numerical solution determines the time taken for a droplet starting with $\xright^0/L=0.7$ to reach $\xright/L=1$ for different values of $\hat{V}$ and $\bendability$ i.e.~we may write
\begin{equation}
\frac{t_{0.7}}{\tau_c}=f(\nu,\hat{V}).
\end{equation} The scaling law \eqref{E:scaling} shows that $f(\nu,\hat{V})\sim (\bendability\hat{V})^{-1}$ in the limit $\hat{V}\ll1$. The numerically-determined values of $t_{0.7}/\tau_c$ are plotted in fig.~\ref{fig:fig3_collapse} for several  values of $\hat{V}$ (spanning the experimentally realized range). These results suggest that some of the discrepancy between experiments and the scaling prediction (\ref{E:small_V_scaling}) are accounted for by the finite value of $\hat{V}$. The neglect of some physical aspects may also result in deviation of experimental results from the numerical solutions; surface defects, the presence of gravity and surface tension acting along the sides will, for example, influence the dynamics. Whilst we expect these to be relatively unimportant~\cite{endnote47}, they will introduce `noise' into experimental results not accounted for by the model.

\begin{figure}[h]
\centering
\includegraphics[width=0.95\columnwidth]{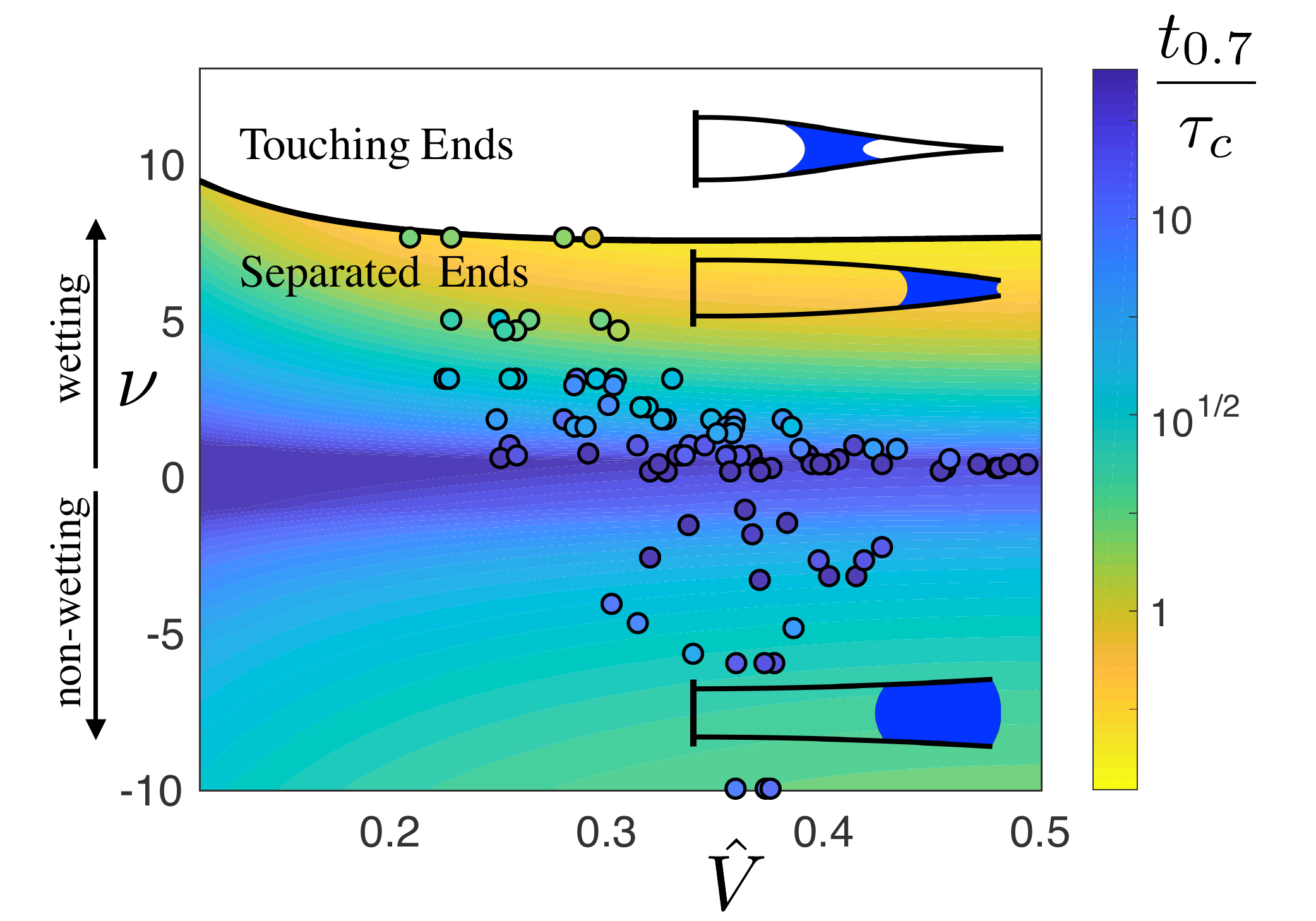}
\caption{\label{fig:fig4_numerics} Influence of  dimensionless droplet volume, $\hat{V}$, and channel bendability, $\bendability$,  on the time taken to traverse the final $30\%$ of the channel, $t_{0.7}/ \tau_{c}$. Numerical results are shown by varying color, while the bold black curve indicates parameter values for which the edges of the channel touch during the motion, trapping the droplet. (Note that the position of this curve depends on the initial condition; here $\xright^0/L = 0.6$.) Positive bendability, $\bendability>0$, corresponds to wetting drops, while $\bendability<0$ corresponds to non-wetting drops; when  $\bendability = 0$, the channel is effectively rigid and the droplet  remains stationary. Schematics illustrate typical configurations, and filled circles correspond to the experimental data presented in figures~\ref{fig:fig2_experiment} and \ref{fig:fig3_collapse}; the outliers with $\nu < 0$  are in the slow non-wetting regime.}
\end{figure}

Numerical solutions of the dimensionless version of \eqref{E:NumSolEq}--\eqref{E:NumSolKBC} yield the values of $f(\nu,\hat{V})$, which are shown in a color-map in fig.~\ref{fig:fig4_numerics} with, schematics of the deformed channel shape.  This demonstrates that, for fixed relative droplet volume $\hat{V}$, the time $t_{0.7}$ decreases as the absolute bendability $| \bendability |$ increases (e.g.~by decreasing the wall thickness $b$, or  Young's modulus $E$). However, this is to be weighed against the possibility of the edges of the walls touching and trapping wetting drops indefinitely (see upper curve in fig.~\ref{fig:fig4_numerics}).

 In this Letter, we have shown that a drop placed into a channel with deformable, but initially parallel, walls creates  its own tapered channel, driving itself towards the free end, independent of the droplet wettability. We suggest that this universality of  motion may find application in self-cleaning surfaces able to remove macroscopic contaminants~\cite{Neinhuis97AnnBot}. In particular, surfaces are often textured at a microscopic  scale to reduce adhesion and increase droplet mobility~\cite{Bico99EPL,Quere05RepProgPhys}. However, these properties can be impaired if  liquid impregnates  the texture~\cite{Cheng05ApplPhysLett}. Tapering the texture has been suggested to reduce the internal fogging of some surfaces~\cite{Mouterde17NMat}, effectively expelling the soiling droplets automatically, but only works if these droplets are themselves non-wetting. A similar role has been suggested for  the hairy coating on the legs of water-walking arthropods such as  \textit{Gerris Regimis} ~\cite{Wang15PNAS}. Here we have shown that under bendotaxis  both wetting and non-wetting drops  move to the free end of a rectangular channel, where they might naturally evaporate, be knocked off or even jump from the surface~\cite{Wisdom13PNAS}. Rapid motion occurs for large values of the bendability, at the risk of trapping wetting droplets (fig.~\ref{fig:fig4_numerics}).
  
There remain many  features of the system (including contact angle hysteresis  three-dimensional geometry, and the behavior of the droplet at the end of the channel) that might complicate the picture of bendotaxis presented here. However, these complications may also provide further opportunities for passive droplet control with more sensitivity: for example, by tapering the undeformed channels slightly, we expect there would be a range of values of the bendability for which droplets would actually move towards the clamped end.

\begin{acknowledgments}

This research was supported by the European Research Council under the European Horizon 2020 Programme, ERC Grant Agreement no. 637334 (DV) and the John Fell Fund, Grant no. BKD00020 (FB).
\end{acknowledgments}

%\bibliography{PRL_bib}

\end{document}

\title{Supplementary Material for\\  ``Wettability-independent droplet transport by \emph{Bendotaxis}''\\
by Bradley, Box, Hewitt \& Vella}

\maketitle
\graphicspath{{./Paper_figures/}}

This supplementary information gives further details on the experimental setup and mathematical model referred to in the main text. In \S \ref{S:experiment} we provide details of our experimental protocols, including fabrication of SLIPS, wettability properties of the channel and a comparison of glass treatments. In \S \ref{S:model} we state the mathematical model in full and discuss in greater depth the method for its numerical solution. We also provide justification for the neglect of the initial transients and discuss the role of gravity. This section also includes a brief asymptotic analysis of the problem for the case in which the drop has small volume (relative to that of the channel).

\section{The experimental procedure}\label{S:experiment}
\subsection{Channel Geometry}
\begin{figure}
\centering
\includegraphics[scale=0.6]{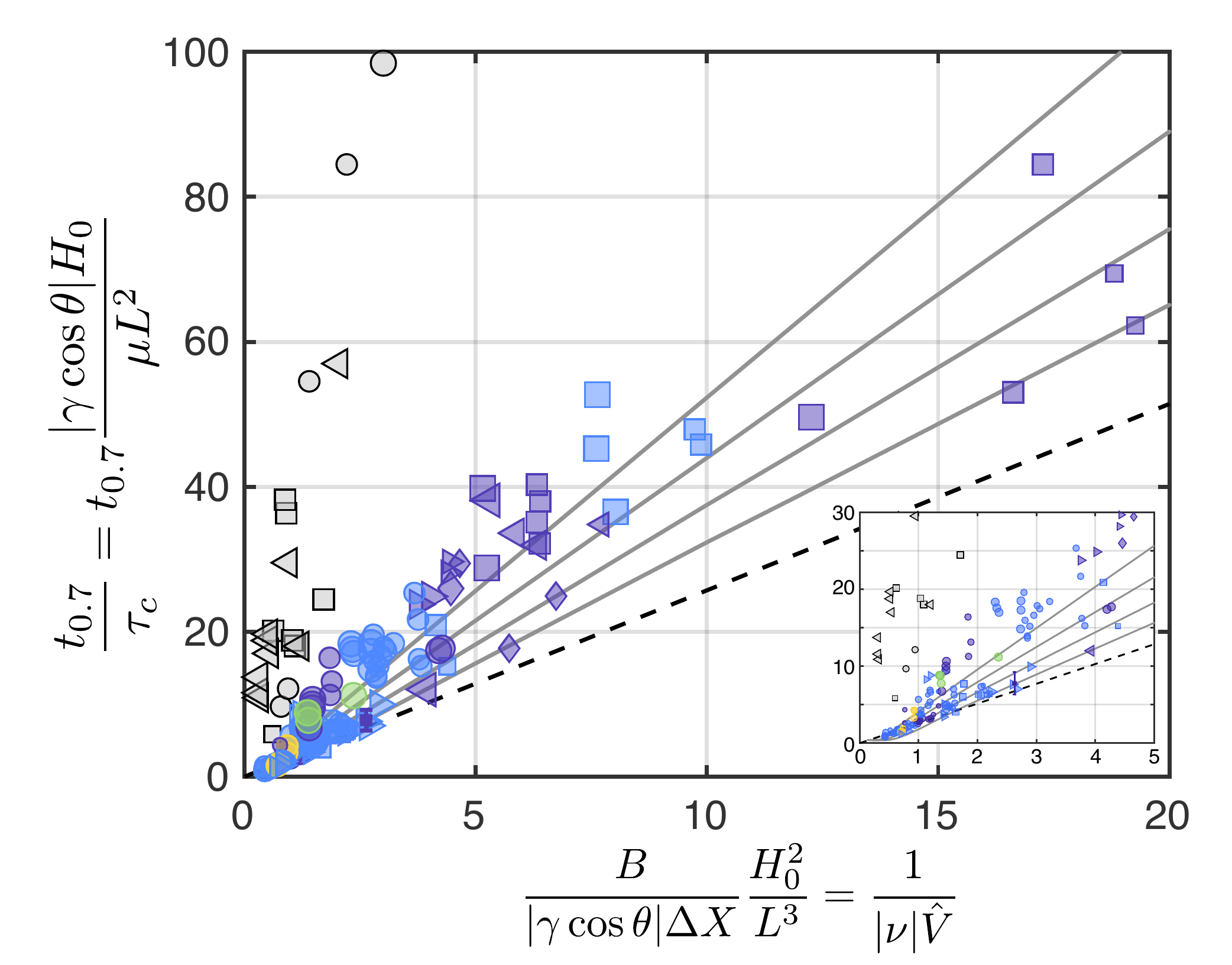}
\caption{Dimensionless experimental data plotted on linear axes. Shape, colour and size of markers encode $2H_0$, $\mu$ and $\hat{V}$ as in fig.~3 of the main text; similarly, the solid lines correspond to solutions of the full mathematical model (described in detail in \S~\ref{S:model}) for dimensionless volumes $\hat{V}=0.5,0.4,0.3$ and $0.2$ (decreasing towards the dashed line, which shows the behaviour in the limit $\hat{V}\ll 1$). Inset: as main figure, zoomed around the origin.}\label{fig:linlin_fig3}
\end{figure}

Borosilicate glass coverslips (thickness numbers 1, 1.5, 2 and 3 from Agar Scientific, UK) were first cut into strips of width $w = 5 \pm 0.5 \mathrm{~mm}$; their thickness, $160\mathrm{~\mu m}\leq b \leq 310\mathrm{~\mu m}~(\pm5\mathrm{~\mu m})$, was measured using a micrometer (Newport, USA). The glass was then treated (as discussed in the next paragraph), but treatment does not significantly affect the thickness of the strips. The strip bending stiffness is therefore calculated as $B = Eb^3/12$~\cite{Howell08} (the Poisson's ratio of glass does not appear in $B$ since the strip is narrow~\cite{audoly10}). We use  a  value of the Young's Modulus, $E = 63\mathrm{~GPa}$, from the literature~\cite{SchottGlassData}. The channel was fabricated by clamping two such strips either side of a rigid glass separator of thickness $310\mathrm{~\mu m}\leq 2H_0 \leq 630 \mathrm{ ~\mu m} ~(\pm5\mathrm{ ~\mu m})$, measured with the same micrometer. 
By varying the channel geometry within these ranges (as well as drop volume), we were able to achieve variations of almost two orders of magnitude in our control parameter $\nu \hat{V}$ (see fig.~3 in the main text). We also present the  same data on linear axes in fig.~\ref{fig:linlin_fig3}.

\subsection{Surface Treatments and Liquid Properties}

We explored bendotaxis for both wetting and non-wetting drops and used a different glass treatment and droplet liquid for each case.

To obtain non-wetting conditions, we used droplets of a glycerol/water mix  (70$\%$ water by weight and henceforth referred to as GWM for brevity) in a channel whose walls were treated to make them into SLIPS with silicon oil V5 (Sigma-Aldrich, USA) used as the lubricant. Further details of the SLIPS fabrication technique are given below (in \S\ref{S:slips}).

To obtain wetting conditions, we used Silicon oil drops in either channels with whose walls had been treated to achieve SLIPS (infused with V100 Silicon oil) or channels that were simply pre-wetted by a Silicon oil of the same viscosity as that used for the particular experiment. We found no significant  difference between the two treatments used in the wetting case as discussed in detail in \S \ref{S:prewetvsSLIPS}. 

We emphasize that bendotaxis \emph{can} be observed for both wetting and non-wetting drops in the \textit{same} channel. In the experiments on non-wetting drops reported here, however, we reduced the lubricant viscosity to ensure drop dissipation is the dominant source of dissipation, which requires the drop viscosity to be at least five times that of the infused coating \cite{Keiser17SoftMatter}.

\subsubsection{Slippery Lubricant Infused Porous Surfaces (SLIPS)}\label{S:slips}
To achieve a slippery lubricant infused porous surface, we first sprayed the target surface with a commercial superhydrophobic coating (Glaco Mirror Coat Zero, Soft 99, Japan), which was then left to dry in ambient conditions. To ensure a robust coating, each target surface was sprayed three times; after the first two applications the surface was left to dry for  30~minutes, whilst after the third, the surfaces were left for 24 hours,  allowing the isopropanol in the spray to completely evaporate. This process left the target surface coated with hydrophobic nano-particles. In our experiments, the thickness of the nano-particle layer is negligible in comparison with that of the target surface.

After drying, the target surfaces were dip coated in Silicon oil (Sigma Aldrich, USA). The Silicon oil penetrates the nano-textures, leaving a robust, lubricating coating on the target surface~\cite{Wong11Nature}. Provided lubricant is not displaced, droplets subsequently introduced onto the surface only share an interface with the lubricant; as a result, these droplets are highly mobile (the surface is `slippery'). For all SLIPS coatings, the speed of withdrawal from the bath of lubricant was controlled using a linear actuator (M-229.26S Physik Instrumente, Germany) in conjunction with a motor controller (C-663 Mercury Step Controller,  Physik Instrumente, Germany). Different viscosity Silicon oils were used for Silicon oil droplets (wetting) and GWM droplets (non-wetting), and so, since the thickness of the lubricating layer follows the Landau--Levich scaling \cite{Guan17SoftMatter}, the product of pulling speed and lubricant viscosity should be preserved. For wetting drops, we used V100 Silicon oil as the lubricant, withdrawing at a speed of $100~\mathrm{\mu m~ s^{-1}}$ (which has been reported to leave a lubricated layer of thickness approximately $3\mathrm{~\mu m}$~\cite{Guan17SoftMatter}); for non-wetting drops, we used V5 Silicon oil as the lubricant,  withdrawing at a speed of $2~\mathrm{mm~ s^{-1}}$.

\subsubsection{Wettability Properties}
In the channels with wetting conditions, droplets of Silicon oil of different viscosities (V50, V100, V350 and V500, all supplied by Sigma Aldrich, USA) were used. We used manufacturer provided values for the density, viscosity and surface tension coefficient of these oils. We observed that drops of these liquids completely wet both SLIPS infused with V100 Silicon oil and glass pre-wetted with the same liquid. We therefore use a contact angle $\theta = 0$ in all subsequent analysis and modelling. (Despite this complete wetting, a clear meniscus of the `capillary bridge' formed between the two plates is visible throughout our experiments.)

Droplets of GWM were used in the channels with non-wetting conditions; the kinematic viscosity was measured using a viscometer (Ametek Brookfield, UK) to be $30 \pm 5 \mathrm{~mm^2 ~ s^{-1}}$. The density of the GWM was measured to be $1.196 \mathrm{~kg ~m^{-3}}$ using a densiometer (DMA 35, Anton Paar GmbH, Austria), from which we calculate a dynamic viscosity of $\mu = 35.9 \mathrm{~m Pa ~s}$. The surface tension coefficient of the GWM was measured  using  the pendant drop method  to be $\gamma=67\mathrm{~mN\,m^{-1}}$, in agreement with previously reported values  \cite{Takamura2012}.

We measured the equilibrium contact angle of a 15$\mathrm{~\mu L}$ droplet  of GWM on a single, horizontal SLIPS infused with V5 silicon oil, i.e.~not within a channel. We analysed images taken with a microscope using the ImageJ contact angle plug-in~\cite{Schneider12NMeth}. This software performs an elliptical fit to the droplet shape and calculates the corresponding contact angles; we measured a value of $\theta_e = 102 \pm 1\si{\degree}$, consistent across a range of images. Errors in this calculation may be significant, since the fit does not account for the lubricant skirt, which has a similar refractive index to the GWM. Further, it has been reported~\cite{Schellenberger15SoftMatter} that microscopic contact angles on SLIPS can vary significantly from those obtained with the elliptic fit technique, while the use of an image from a single SLIPS, rather than a channel makes the role of the skirt difficult to quantify. 

\begin{figure}[h]
\includegraphics[scale=0.55]{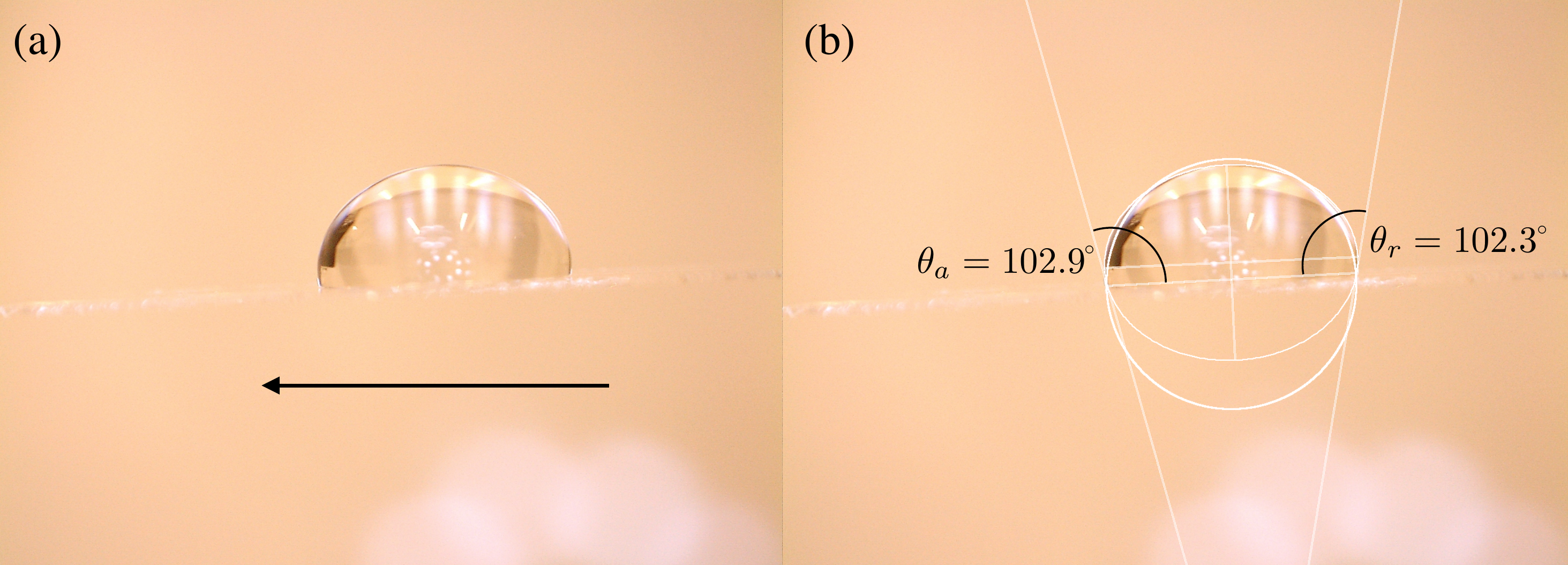}
\caption{\label{fig:contact_angle}Contact angle measurements of a Glycerol/Water Mix (GWM) droplet on a SLIPS infused with V5 silicon oil. (a) Image taken as a $15\mathrm{~\mu L}$ droplet slides down the surface, which is inclined by approximately 1$\si{\degree}$ to the horizontal. The black arrow indicates the direction of motion. (b) The same image with  an elliptical fit applied to the drop using the contact angle plug-in in ImageJ. The angles measured in this image are $\theta_a = 102.9\si{\degree}$ and $\theta_r =  102.3\si{\degree}$. }

\end{figure}

For the advancing and receding contact angles, we used the same method, albeit with images of a $15\mathrm{~\mu L}$ droplet  sliding down a SLIPS inclined by $\approx 1 \si{\degree}$ to the horizontal. (Whilst it is typical to determine dynamic contact angles by measuring angles just as the drop begins to move on an inclined plate~\cite{Yuan2013contact}, we measured angles during the motion down the surface, as it is these angles that are used in our model).
Again, we took many measurements, with mean advancing contact angle of $\theta_a = 101.8\pm1.7\si{\degree}$ (error represents one standard deviation of measured values) and receding contact angle of $\theta_r = 100.2\pm2.2\si{\degree}$ using this method (one such measurement is shown in fig.~\ref{fig:contact_angle} with $\theta_a = 102.9\si{\degree}$, $\theta_r = 102.3\si{\degree}$). We note that in each measurement the contact angles were found to satisfy  $\theta_r < \theta_a$, as expected.

Previous studies (for example~\cite{Onda1996Langmuir}) have successfully used similar nano-scale roughness to achieve very high contact angles; it is desirable in studying bendotaxis (for non-wetting channels) to have as large a contact angle as possible as this minimizes the influence of both gravity (see \S\ref{S:gravity}) and the line force from surface tension acting at the menisci.  Unsuccessful attempts were made to perform experiments on slippery superhydrophobic surfaces (with a contact angle with water above 150$\si{\degree}$ and negligible roll angle). These attempts were unsuccessful because of difficulties associated with introducing the drop into the channel: the droplet would be immediately ejected from the channel along one of the transverse edges or resist entry entirely (even when hydrophobized syringe tips were used).

\subsection{Performing the Experiment}
A droplet was introduced into the channel using a micropipette (Better Equipped, UK) (for the thinnest channels, we attached a syringe tip of diameter $250 \mathrm{~\mu m}$ to the micropipette tip to reduce the risk of damaging the clamped region or contaminating the channel). During this insertion of a droplet, the two channel walls were held separated by a distance $2H_0$  at (what would become) the free end. The droplet moved slowly towards the centre of the channel before the free ends were released (again because of bendotaxis).  However, the ends were released after $\lesssim 10\mathrm{~s}$, which is much shorter than the time scale of motion within the doubly clamped channel (see \S \ref{S:intial_condition}) so the motion in this `doubly clamped' state is ignored.

A digital camera (Nikon D700), mounted above the experiment, captured an image from above every $1~\mathrm{s}$ (with images having a resolution of 1920$\times$1080 pixels, corresponding to a typical spatial resolution of 0.03~mm/pixel). Since the menisci were not perfectly two-dimensional when viewed from above (see fig.~1 of the main text), we defined the values of $\xleft$ and $\xright$ to be the extent of the menisci measured along the centreline of the channel; these positions were obtained using the Canny edge detection algorithm, implemented in \textsc{Matlab}~\cite{Canny86}. The same edge-detection technique was employed to determine the outline of the channel walls viewed from the side, which is also shown in fig.~1 of the main text.

\subsection{SLIPS vs. Prewetting}\label{S:prewetvsSLIPS}

To obtain wetting conditions with low contact angle hysteresis, we used both SLIPS and smooth glass pre-wetted with the same Silicon oil. Both treatments also reduce the experiments' sensitivity to glass defects (which were effectively `smoothed over' by the lubricating or pre-wetting layer).

While the qualitative effect of SLIPS and pre-wetting layers are similar, it is not clear that their dynamic behaviour will be quantitatively similar. We therefore performed an additional experiment to test whether there was any difference between pre-wetted and SLIPS channels (i.e. to assess the effect of application of the hydrophobic spray). In this experiment we conducted 45 repetitions of bendotaxis with a (nominally) fixed channel geometry ($L = 27\mathrm{~mm}, 2H_0 = 430\mathrm{~\mu m}$, $w = 5\mathrm{~mm} $, $b = 180~\mathrm{\mu m}$) and varying whether the channel walls were SLIPS or pre-wetted. We used droplets of the same liquid (Silicon oil V50) and volume $V = 15\mathrm{~\mu L}$; small variations in channel wall separation in different experiments meant that the relative volume $\hat{V}$ varied in the interval $0.29 \leq \hat{V} \leq 0.35$, with a mean value $\hat{V}_{\text{mean}} = 0.32$ and standard deviation of 0.02.

\begin{figure}[h]
\includegraphics[scale=0.7]{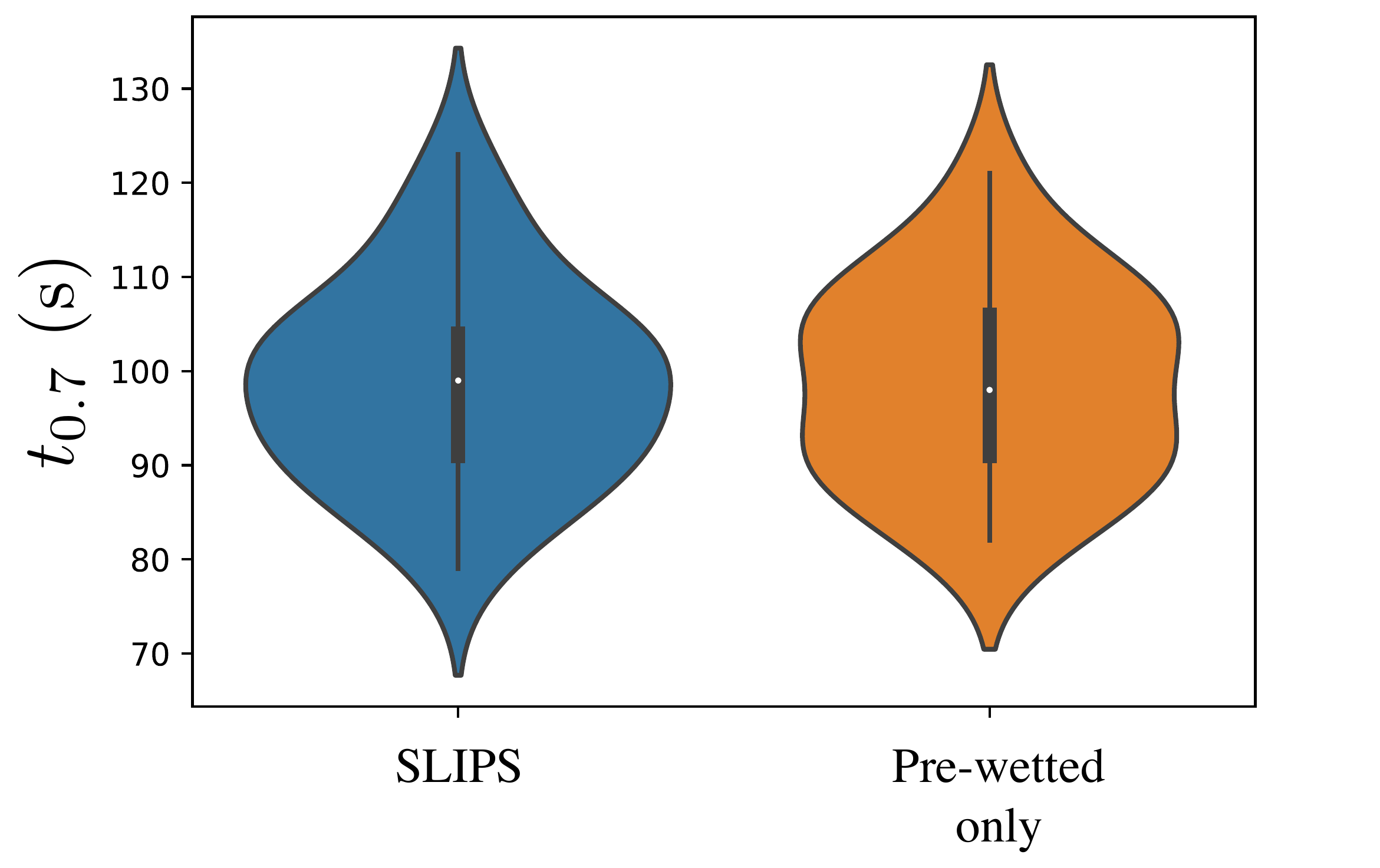}
\caption{\label{fig:SLIPSvsPrewetting}Violin plot of experimentally obtained values of $t_{0.7}$ for channels prepared with walls treated to achieve a SLIPS (left) and a pre-wetted surface (right). The data represent the results of performing the bendotaxis experiment many times (25 for SLIPS and 20 for pre-wetted) using channel  geometry $L = 27\mathrm{~mm}, 2H_0 = 430\mathrm{~\mu m}$, $w = 5\mathrm{~mm} $, $b = 180~\mathrm{\mu m}$ and V50 Silicon oil droplets of volume $V=15\mathrm{~\mu L}$. The mean values are $\bar{t}_{0.7} = 98.85$~s and  $\bar{t}_{0.7} = 98.89$~s in the SLIPS and pre-wetted cases, respectively, with standard deviations of 10.8~s and 10.6~s.}
\end{figure}

We carried out 25 tests with SLIPS channels and 20 with pre-wetted channels; fig.~\ref{fig:SLIPSvsPrewetting} shows violin plots of the values of $t_{0.7}$ for both cases; the kernel density estimates have very similar shapes about an identical mean $\bar{t}_{0.7} = 98.9 \mathrm{~s}$. We conclude that there is no significant difference in $t_{0.7}$ between the two treatments (SLIPS and pre-wetting) and therefore do not differentiate between them in the figures and analysis in the main text.

\section{The mathematical model}\label{S:model}

\subsection{Dimensional Mathematical Model}\label{S:dim_math_model}

We use linear beam theory to model the deflection (along the channel) of the strips in response to hydrodynamic pressure. Strip inertia is neglected, since the inertial time scale of an elastic strip is obtained by balancing inertia and bending stiffness, $\tauinert \sim (\rho_s b L^4/B)^{1/2}=(12\rho_s L^4/Eb^2)^{1/2}\sim 1\mathrm{~ms} \ll \tau_c$, the capillary time scale. Tension within the strip may also be neglected since the end is free and the relative size of the jump induced by the meniscus is expected to scale with $L^2 \gamma /B\sim 10^{-6} \ll 1$. For the moment, we shall also neglect gravitational forces,  postponing a more formal justification to \S\ref{S:gravity}.

 Assuming up-down symmetry, only one strip need be considered; the separation of the channel walls, $2h(x,t)$, satisfies the quasi-static (zero inertia) beam equation
\begin{equation}\label{E:beam_eq_dim}
B\frac{\partial^4 h}{\partial x^4} = q(x,t)\,,
\end{equation}
where $q(x,t)$ is the  pressure applied by the droplet, so that
\begin{equation}
 q(x,t) = \left\{\begin{array}{lr}
       0 \qquad \qquad  & \text{for } 0<x<\xleft(t), \\
        p_{\text{drop}}(x,t)  & \text{for}\ \xleft(t)\leq x\leq \xright(t),\\
        0 &   \text{for } \xright(t) \leq x\leq L.
        \end{array}\right.
\end{equation}
 Thusfar, the droplet pressure $p_{\text{drop}}(x,t)$ is undetermined; this pressure is determined by the coupling between the flow in the liquid and the deflection of the walls. We apply lubrication theory~\cite{Leal07} to model the behaviour of the drop with the hydrostatic pressure contribution neglected (the drop Bond number scales as $\rho_l g H_0^2/\gamma \ll 1$). In this framework, the droplet pressure satisfies Reynolds' equation
\begin{equation}\label{E:RLE_dim}
\frac{\partial h}{\partial t} = \frac{\partial}{\partial x}\left(\frac{h^3}{3\mu} \frac{\partial p_{\text{drop}}}{\partial x}\right), \qquad \xleft(t) < x < \xright(t).
\end{equation}
Combining \eqref{E:beam_eq_dim} and \eqref{E:RLE_dim} yields the equation for the evolution of the wall separation in $\xleft(t) < x < \xright(t)$:
\begin{equation}\label{E:beam_wet_dim}
\frac{\partial h}{\partial t}  = \frac{B}{3 \mu}\frac{\partial}{\partial x}\left[h^3 \frac{\partial^5 h}{\partial x^5}\right]\,,
\end{equation}
whilst the channel wall separation within the regions  not in contact with the droplet satisfy 
\begin{equation}\label{E:beam_dry_dim}
\frac{\partial^4 h}{\partial x^4} = 0, \qquad \text{for} \quad 0 < x < \xleft(t), \quad \text{and} \quad \xright(t) < x < L.
\end{equation}
The rate of evaporation is negligible on the time scale of motion (as justified in \S\ref{S:evaporation}); the flux through the menisci must therefore balance that caused by the motion, giving the kinematic conditions
\begin{equation}\label{E:kinematic_dim}
\frac{\mathrm{d}x_m}{\mathrm{d}t} = -\frac{B h^2}{3\mu}\left.\frac{\partial h^5}{\partial x^5}\right|_{x = x_m}, \qquad\mathrm{for}\quad x_m = \xleft, \xright.
\end{equation}
Note that these conditions automatically ensure that the volume of the droplet is conserved throughout the motion described by \eqref{E:beam_wet_dim}.
The problem \eqref{E:beam_wet_dim}-\eqref{E:kinematic_dim} is closed by specifying boundary conditions at $x = 0$ and $x = L$, matching conditions across the menisci and an initial condition; the strips are clamped at their base,
\begin{align}\label{E:clampedBC_dim}
h(0,t) &= H_0, & \left.\frac{\partial h}{\partial x}\right|_{x = 0} &= 0, 
\end{align}
and free at $x =L$,
\begin{align}\label{E:freeBC_dim}
\left.\frac{\partial^2 h}{\partial x^2}\right|_{x = L} &= 0, & \left.\frac{\partial^3 h}{\partial x^3}\right|_{x = L} &= 0. 
\end{align}
Channel wall separation, slope, moment and shear are continuous across the menisci at $x = x_\pm$ (we neglect any line force at the menisci, since the ratio of line force to drop pressure scales as $H_0\tan \theta/L $~\cite{Taroni12JFM}; in our experiments, $H_0\tan \theta/L < 0.05$ even in the non-wetting conditions, where $\theta$ is reasonably close to $\pi/2$):

\begin{equation}\label{E:matching_cond_dim}
\left[h\right]_{x_{\pm}^-}^{x_{\pm}^+} = \left[\frac{\partial h}{\partial x}\right]_{x_{\pm}^-}^{x_{\pm}^+}  =  \left[\frac{\partial^2 h}{\partial x^2}\right]_{x_{\pm}^-}^{x_{\pm}^+} = \left[\frac{\partial^3 h}{\partial x^3}\right]_{x_{\pm}^-}^{x_{\pm}^+} = 0,
\end{equation}
where $[f]_-^+ = f(\zeta_+) - f(\zeta_-)$ denotes the jump in the value of a function across a point. The motion is driven by the Laplace pressure immediately beneath each meniscus; this known pressure provides boundary conditions on the strip shape via an imposed pressure boundary condition at each meniscus, which read:
\begin{equation}\label{E:pressure_dim}
B\left[ \frac{\partial^4 h}{\partial x^4}\right]_{x_\pm^-}^{x_\pm^+} = \left.\frac{-\gamma \cos \theta}{h}\right|_{x=x_\pm}.
\end{equation}
Finally, the initial conditions are
\begin{equation}\label{E:IC_dim}
h(x, t= 0) = H_0, \qquad \xleft(0) = x_-^0, \qquad \xright(0) = x_+^0.
\end{equation}
(The perturbation to the initial condition resulting from bending in the initial `doubly-clamped' regime is ignored; see \S\ref{S:intial_condition} for a justification of this.)

\subsection{The Role of Gravity and Evaporation}

\subsubsection{Gravity}\label{S:gravity}
In this subsection, we consider the pressure gradient induced by gravity, relative to that caused by elastocapillary effects. As discussed in the main text, for a droplet of small relative volume $\hat{V}$, the elastocapillary pressure gradient  is $p_x^{\text{ec}} \sim \gamma^2 \cos^2 \theta \ L^2 \Delta X / (BH_0^3)$. (Note that accounting for larger $\hat{V}$ increases the elastocapillary pressure gradient.)

The pressure gradient due to gravity arises from a deflection of the whole strip under its own weight (net force $F_g^{\text{beam}} \sim \rho_s g b Lw\approx 6\times10^{-4}\mathrm{kg}$) and that of the drop (net force $F_g^{\text{drop}} \sim \rho_d g V\approx2\times10^{-4}\mathrm{~kg}$). By considering a cantilever beam weighted down by forces $F_g^{\text{beam}}$ and $F_g^{\text{drop}}$, we find the induced angle is $\alpha_g \sim (\rho_ s b + \rho_d H_0 \hat{V}) g L^3 /B$ (relative to the horizontal). The associated pressure gradient is $p_x^{\text{g}} \sim \rho_d   g   \alpha_g \sim (\rho_ s b + \rho_d H_0 \hat{V}) \rho_d     g^2  L^3/B$. The ratio of these two pressure gradients is $p_x^{\text{g}}/p_x^{\text{ec}} \sim (\rho_ s b + \rho_d H_0 \hat{V}) L H_0^3 /  (\rho_d \Delta X \ell_c^4)$, where $\ell_c = \sqrt{\gamma |\cos \theta|/(\rho_d g)}$ is the capillary length ($\approx  1.5\mathrm{~ mm}$ for Silicon oil). For our channels the ratio  $p_x^{\text{g}}/p_x^{\text{ec}}< 5\times 10^{-2}$ and so we neglect the bending of the plates under their own weight (or that of the droplet) as a driving mechanism for motion. 

\subsubsection{Evaporation}\label{S:evaporation}
A full description of the evaporation dynamics of droplets in channels is far beyond the scope of the present study. Therefore, to assess the time-scale of evaporation in our experiments, we consider the evaporation of a sessile droplet of similar size. (Note, however, that we expect evaporation of a confined drop to be slower than this --- the surface area exposed to the air is significantly lower because of the channel walls.) The timescale over which the droplet evaporates is $\tau_{\text{evap}} \sim \rho V / R_d D c_v$, where $R_d$ is the drop radius, $D$ is the gas phase diffusivity, and $c_v$ the vapour concentration at saturation \cite{Hu2002JPhysChemB}. Using the values $D = 17~\mathrm{mm^2~ s^{-1}}$, $c_v =19 \times 10^{-3}~\mathrm{kg~m^{-3}}$ for water (also from ref.~\cite{Hu2002JPhysChemB}) and typical  values for the drop radius and volume in our experiments, we find $\tau_{\text{evap}} \approx 5$ hours; this is much longer than any experiment took to complete, and therefore evaporation can be safely neglected. Furthermore, we note that our assertion that this time scale is a lower bound for the time scale of evaporation agrees with our experimental observations that drops left in channels overnight reduced in size but had not completely evaporated by the next morning.

\subsection{Dimensionless model}
With the non-dimensionalization described in  the main text (i.e.~$x = L\x, h = H_0 \h, t = \tau_c \tc $) we have that the separation of the channel walls, $2\h(\x,\tc)$, satisfies
\begin{align}
0& = \frac{\partial^4 \h}{\partial \x^4}\;, & &0 < \x <\xl(t)\;,\label{E:Eq_left} \\
\frac{\partial \h}{\partial \tc} &= \frac{1}{3|\nu|}\frac{\partial}{\partial \x}\left[\h^3 \frac{\partial^5 \h}{\partial \x^5}\right]\;, & &\xl(t)<\x<\xr(t)\;, \label{E:Eq_wetted}\\
0& =\frac{\partial^4 \h}{\partial \x^4}\;,  & &\xr(t)< \x < 1,\label{E:Eq_right}
\end{align}
with the bendability $\nu = \gamma \cos \theta L^4 /(BH_0^2)$ as before.
The  boundary and matching conditions~\eqref{E:clampedBC_dim}-\eqref{E:matching_cond_dim} become
\begin{align}\label{E:BC_base}
\left.\h\right|_{\x=0} &= 1, & \left.\frac{\partial \h}{\partial \x}\right|_{\x=0} &= 0, & \left.\frac{\partial^2 \h}{\partial \x^2}\right|_{\x=1} &=0, &  \left.\frac{\partial^3 \h}{\partial \x^3}\right|_{\x=1} &= 0,
\end{align}
and
\begin{align}\label{E:BC_continuity}
\left[\h\right]^{\x_\pm^+}_{\x_\pm^-} &= 0\;, & \left[\frac{\partial \h}{\partial \x}\right]^{\x_\pm^+}_{\x_\pm^-} &= 0\;, &  \left[\frac{\partial^2 \h}{\partial \x^2}\right]^{\x_\pm^+}_{\x_\pm^-} &= 0\;, &  \left[\frac{\partial^3 \h}{\partial \x^3}\right]^{\x_\pm^+}_{\x_\pm^-} &= 0\;.
\end{align}
The pressure condition~\eqref{E:pressure_dim} becomes
\begin{align}\label{E:BC_pressure}
\left.\frac{\partial^4 \h}{\partial \x^4}\right|_{\x = \x_{\pm}} &= \frac{-\nu }{\h(\x_{\pm},\tc) }, 
\end{align}
and the meniscus positions evolve according to
\begin{equation}\label{E:KBC}
\frac{\mathrm{d}\x_{\pm}}{\mathrm{d}\tc} = -\left. \frac{\h^2}{3|\nu|} \frac{\partial^5 \h}{\partial \x^5}\right|_{\x = \x_{\pm}}.
\end{equation}
The initial condition is
\begin{align}\label{E:IC}
\h(\x, \tc= 0) &= 1, & \xr(0) &= \xr^0, & \xl(0) &= \xl^0,
\end{align}
which specifies the dimensionless drop volume $\hat{V} = \xr^0 - \xl^0$.

\subsection{Numerical solution}

To obtain numerical solutions of the full problem \eqref{E:Eq_left}--\eqref{E:IC}, we first transform the problem to one defined only on the drop region $\xl < \x < \xr$. This is possible because the solutions in $0 < \x < \xl$ and $\xr < \x < 1$ may be determined analytically and used to give explicit expressions for the boundary conditions at the menisci.

In more detail, the channel wall separations in the regions outside the drop are simply cubic functions of $\x$ with coefficients dependent on the strip shape within it -- they vary with time only through the meniscus positions. Explicitly, these solutions are
\begin{equation} \label{E:non-wetted}
 \h(\x,\tc) = \left\{\begin{array}{lr}
       \left(\frac{\x}{\xl}\right)^3\left.\left[\xl \frac{\partial \h}{\partial \x}\ - 2\h + 2\right]\right|_{\xl} + \left(\frac{\x}{\xl}\right)^2\left.\left[3\h - \xl \frac{\partial \h}{\partial \x} -3\right]\right|_{\xl} + 1 & \text{for } 0 < \x < \xl(\tc),\\
  \left( \x - \xr\right)    \left.\frac{\partial \h}{\partial \x}\right|_{\xr}+ \left. \h \right|_{\xr} & \text{for } \xr(\tc) < \x < 1.
        \end{array}\right. 
\end{equation}
In the drop region, the governing equation
\begin{align}\label{E:numerics_eq}
\frac{\partial \h}{\partial \tc} &= \frac{1}{3|\nu|}\frac{\partial}{\partial \x}\left[\h^3 \frac{\partial^5 \h}{\partial \x^5}\right]\;, & &\xl<\x<\xr\;,
\end{align}
still holds, but the boundary conditions may be written explicitly as:
\begin{equation}\label{E:numeric_BC-}
\frac{\partial^2 \h}{\partial \x^2} = \frac{2}{\xl^2}\left(2 \xl \frac{\partial \h}{\partial \x} - 3\h  +3\right)\;, \quad
\frac{\partial^3 \h}{\partial \x^3} = \frac{6}{\xl^3}\left(\xl \frac{\partial \h}{\partial \x} - 2\h +2\right)\;, \quad
\frac{\partial^4 \h}{\partial \x^4}= \frac{-\nu }{\h} \;,\quad \mathrm{at}\quad x = \xl\;,
\end{equation} and 
\begin{equation}\label{E:numeric_BC+}
\frac{\partial^2 \h}{\partial \x^2} = 0\;, \quad
\frac{\partial^3 \h}{\partial \x^3} = 0\;, \quad
\frac{\partial^4 \h}{\partial \x^4}= \frac{-\nu }{\h}\;,\quad \mathrm{at}\quad \x =\xr.
\end{equation}

Once the problem \eqref{E:numerics_eq}--\eqref{E:numeric_BC+} is solved, subject to~\eqref{E:KBC}--\eqref{E:IC}, the behaviour of the whole system can be constructed using~\eqref{E:non-wetted}, together with the continuity conditions~\eqref{E:BC_continuity}. The main focus then is to solve \eqref{E:numerics_eq}--\eqref{E:numeric_BC+}; this is made easier by transforming the drop region $\xl(\tc)<\x<\xr(\tc)$, which evolves in time, onto a fixed domain. We let
\begin{equation}
z = \frac{\x - \xl(\tc)}{\xr(\tc) - \xl(\tc)}\;, \qquad  0 < z < 1.
\end{equation}
Since $z$ is time dependent, this transformation results in additional advective terms in~\eqref{E:numerics_eq}. However, by letting
\begin{equation}
U(z,\tc) = (\xr -\xl) \h(\x,\tc)
\end{equation}
the transformed version of \eqref{E:numerics_eq} may be written in the flux-conservative form as
\begin{equation}
\label{eqn:FluxCons}
\frac{\partial U}{\partial \tc} + \frac{\partial Q}{\partial z} = 0\;,
\end{equation}
where the flux $Q$ is
\begin{equation}\label{E:Qflux}
Q =  -\frac{U}{ (\xr - \xl)}\left[\frac{U^2}{3|\bendability|(\xr - \xl)^8}\frac{\partial^5 U}{\partial z^5} +(1-z) \frac{\upd \xl}{\upd \tc}+ z \frac{\upd \xr}{\upd \tc}\right].
\end{equation}
The problem \eqref{eqn:FluxCons}  is solved by discretizing in space: the $z$-domain is divided into a grid of $n$ cells of equal length $\Delta z=1/n$ with cell centres $z_{j} = (j - \tfrac{1}{2})\Delta z$ for $j = 1, \dots, n$, and edges at $z_{j+1/2} = j \Delta  z$ for $j = 0,\dots,n$.  (Here,  numerical results are presented with $n = 100$.) $U(z,\tc)$ is approximated at the centres by $U_j = U(z_j, \tc)$. Three ghost points are introduced at each end of the domain (corresponding to cell centres indexed by $j = -2,-1,0$ and  $j = n+1,n + 2, n+3$); $U$ is also approximated at these points to implement the (transformed versions of) boundary conditions~\eqref{E:numeric_BC-} and~\eqref{E:numeric_BC+}. The flux $Q$ is approximated at the edges of each cell by $Q_{j + 1/2} = Q(z_{j+1/2},\tc)$; the values of $Q_{j+1/2}$ are obtained using second order centred finite differences of $U_{j},~ j = -2, \dots, n+3$, with $U(z_{j+1/2},\tc)=(U_j+U_{j+1})/2$ and the averaging method of ref.~\cite{Zhornitskaya00} applied for the $U^3$.  This finite difference discretization results in a system of $n$ ODEs, namely
\begin{equation}\label{E:NumEq1}
\frac{\upd U_j}{\upd \tc} = -\frac{Q_{j+\tfrac{1}{2}} - Q_{j-\tfrac{1}{2}}}{\Delta z}\;, \qquad j = 1,\dots,n.
\end{equation}
These are coupled to the kinematic conditions
\begin{align}\label{E:NumEq2}
\frac{\mathrm{d} \xl}{\mathrm{d}  \tc} & = -\left. \frac{U^2}{3 |\bendability| (\xr - \xl)^8}\frac{\partial^5 U}{\partial z^5}\right|_{z = 0}\;, & \frac{\mathrm{d} \xr}{\mathrm{d}  \tc} & = -\left. \frac{U^2}{3 |\bendability| (\xr - \xl)^8}\frac{\partial^5 U}{\partial z^5}\right|_{z = 1},
\end{align}
which are discretized using centered finite differences for the derivatives and a second order, one-sided approximation for the $U^2$ term; by using a one-sided method here, we reduced numerical error in conservation of mass (compared to a centred method). 

The $n+2$ ODEs~\eqref{E:NumEq1}--\eqref{E:NumEq2} are solved using MATLAB's stiff differential equation solver \textsc{ODE15s}. This method exploits the sparsity of the Jacobian, which is calculated using complex step differentiation~\cite{Shampine07TOMS}. Typical computation time is on the order of seconds, but grows rapidly as the dimensionless drop length $\xr - \xl $ approaches zero, owing to the sensitive dependence of the system on this quantity.

\subsection{Asymptotics for small droplets: $\hat{V}\ll 1$}\label{S:smallvhat}
Analytical progress can be made in the regime of small volume droplets, $\hat{V} \ll 1$, even when the bendability $\bendability$ remains order unity (experimentally,  $0.1 < |\bendability| < 10$). In this limit, the droplet does not significantly bend the channel and we therefore assume
\begin{equation}\label{E:small_disp}
\h(\x,\tc) = 1 + \smallpar \f(\x,\tc) + \mathcal{O}\left(\smallpar^2 \right)
\end{equation} 
and anticipate that $\f \sim \mathcal{O}(1)$ (which we verify \textit{a posteriori}).

Note that conservation of  mass then requires the length of the drop to remain small and approximately constant throughout the motion, $\xr - \xl = \hat{V} + \mathcal{O}(\smallpar ^2)$. Therefore, by integrating the governing equation~\eqref{E:Eq_wetted} across the drop, we find that the flux, $\h^3\partial^5 \h /\partial \x^5$, does not vary significantly across it:
\begin{equation}
\left[\h^3 \frac{\partial^5 \h}{\partial \x^5}\right]^{\xr}_{\xl} = \int_{\xl}^{\xr} \  3 |\bendability| \frac{\partial \h}{\partial \tc} \ \mathrm{d}\x=3|\bendability| \int_{\xl}^{\xr}\smallpar \frac{\partial \f}{\partial \tc}~\upd \x = \mathcal{O}\left(\smallpar^2\right).
\end{equation}
As a consequence, the pressure gradient, $ \partial^5 \h /\partial \x^5$, is approximately constant in the drop and, using~\eqref{E:BC_pressure}, we calculate this to be
\begin{equation}\label{E:pressure_gradient}
\frac{\partial^5 \h}{\partial \x^5} \approx \frac{1}{\xr - \xl}\left[\frac{\partial^4 \h}{\partial \x^4}\right]^{\xr}_{\xl}  = \frac{\nu}{\xr - \xl}\left[ \frac{1}{\h(\xl,\tc)} - \frac{1}{\h(\xr,\tc)}\right] = \bendability \hat{V} \left.\frac{\partial \f}{\partial \x}\right|_{\x = \xr} + \mathcal{O}\left(\smallpar^2\right)
\end{equation}

The evolution equation for the meniscus positions, \eqref{E:KBC}, then gives, to leading order,
\begin{equation}\label{E:asy_KBC}
\frac{\mathrm{d} \xr}{\mathrm{d} \tc} =- \frac{\nu}{|\nu|}\frac{\smallpar}{3}\left.\frac{\partial \f}{\partial \x}\right|_{\x = \xr}.
\end{equation}

To proceed, we require an estimate of the slope of the channel walls. Since the drop is small, its effect on the channel shape may be approximated by a point force acting at $\x = \xr$. Jump conditions describing this force are derived by integrating the pressure across the drop: from \eqref{E:pressure_gradient} we see that the pressure gradient within the drop is $\mathcal{O}(\smallpar)$ so that the pressure, $\hat{p} = \partial^4 \h /\partial \x^4$, is approximately constant and given by 
\begin{align}
\hat{p}(\x,\tc) = \hat{p}(\xr,\tc) + \mathcal{O}(\smallpar^2)  = -\bendability + \mathcal{O}\left(\smallpar^2\right)\;, & &\xl < \x < \xr.
\end{align}
By integrating this expression, we calculate the  jump in shear force across the drop:
\begin{equation}
\left[\frac{\partial^3 \h}{\partial \x^3}\right]^{\xr}_{\xl} = -\bendability \smallpar  + \mathcal{O}\left(\smallpar^2\right).
\end{equation}
Since $\partial^3 \h/ \partial \x^3= 0$ at $\x = \xr$, it follows that
\begin{align}\label{E:asy_h_third}
\h_{\x\x\x}(\x,\tc) =  \mathcal{O}\left(\smallpar\right)\;, & & &\xl < \x < \xr.
\end{align}
By a similar argument, we find
\begin{align}
\left[\h_{\x\x}\right]^{\xr}_{\xl}  &= \mathcal{O}\left(\smallpar^2\right)\;, & \left[\h_{\x}\right]^{\xr}_{\xl}  &= \mathcal{O}\left(\smallpar^3\right).
\end{align}
The leading order `point force problem' for $\f$ is therefore
\begin{align}\label{E:asy_eq1}
\frac{\partial^4 \f}{\partial \x^4} &= 0\;, & &\text{in} \ 0 < \x < \xr\;, \ \xr < \x < 1\;,
\end{align}
with jump conditions
\begin{align}\label{E:asy_eq2}
\left[\frac{\partial^3 \f}{\partial \x^3} \right]_{\xr^-}^{\xr^+} &= - \nu \;, & \left[ \frac{\partial^2 \f}{\partial \x^2} \right]_{\xr^-}^{\xr^+} &= 0\;, & \left[\frac{\partial \f}{\partial \x} \right]_{\xr^-}^{\xr^+} &= 0\;, & \left[ \f \right]_{\xr^-}^{\xr^+} &= 0\;,
\end{align}
and boundary conditions
\begin{align}\label{E:asy_eq3}
\f &= 0 = \frac{\partial \f}{\partial \x}\;, & &\text{at} \quad x= 0\;,\\
\frac{\partial^2 \f}{\partial \x^2}&= 0 =\frac{\partial^3 \f}{\partial \x^3} \;, & &\text{at} \quad x= 1.\label{E:asy_eq4}
\end{align}
The solution to~\eqref{E:asy_eq1}--\eqref{E:asy_eq4} is
\begin{equation}\label{E:solutionf}
   \f(\x,\tc) = \begin{cases}{}
        -\frac{\bendability}{6}(3\xr \x^2 - \x^3) & \text{for }\quad 0 < \x < \xr(\tc)\;, \\
       -\frac{\bendability}{6}\bigl[3\xr^2(\x - \xr) + 2 \xr^3\bigr] & \text{for }\quad \xr(t) < \x < 1.
        \end{cases}
\end{equation}
(Note that this is consistent with~\eqref{E:non-wetted} in the limit $\xl \approx \xr$.)

\begin{figure}
\centering
\includegraphics[width=0.7\columnwidth]{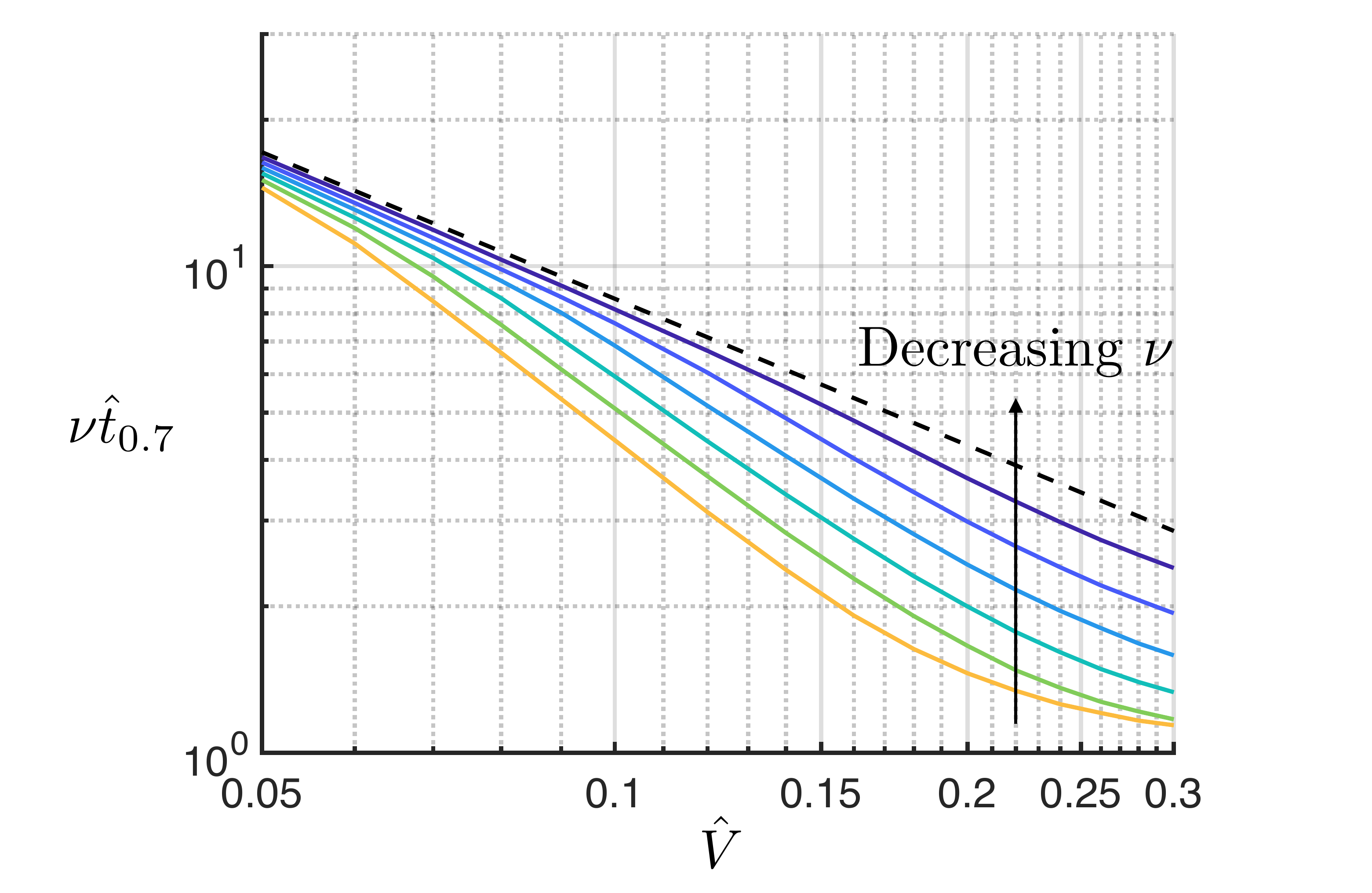}
\caption{Comparison of numerically obtained values of $\nu \hat{t}_{0.7}$ (solid lines) and the asymptotic result~\eqref{E:asy_result} (dashed black line). Data are shown for integer values of $\nu$ in the range $4 \leq \nu \leq 9$, which corresponds to the majority of the experimentally realised parameter range.}
\label{fig:small_v_convergence}
\end{figure}

By inserting~\eqref{E:solutionf} into the reduced evolution equation for the meniscus positions,~\eqref{E:asy_KBC}, we obtain an ODE for the leading order evolution of the meniscus position, expressed in terms of $\hat{V}$ as
\begin{equation}
\frac{\mathrm{d}\xr}{\mathrm{d}\tc} = \frac{|\bendability| \hat{V}\xr^2}{6}.
\end{equation}
This may be solved with the initial condition $\xr(0) =X$ to give
\begin{equation}
\xr(\tc) = \frac{6 X}{6- |\bendability| \smallpar X\tc}.
\end{equation}
In particular, this gives the time taken for the drop to travel from  $\xr = X$ to $\xr = 1$,  $\tc_{X}$, as
\begin{equation}\label{E:asy_result}
\tc_{X} = \frac{6(1-X)}{|\bendability| \smallpar X}.
\end{equation}
By rescaling the variables, the result~(7) from the main text is obtained. In figure~\ref{fig:small_v_convergence}, we plot numerically obtained values of $\bendability \tc_{0.7}$ against $\hat{V}$ for several different values of $\bendability$; we see that these numerically obtained values agree with the asymptotic result~\eqref{E:asy_result} in the limit $\hat{V} \to 0$.

\subsection{The doubly clamped state}\label{S:intial_condition}

As described in \S\ref{S:experiment}, there is a period of time immediately after the drop is inserted into the channel during which the channel ends at $x = L$  are not free, but are in fact held at a distance $2H_0$ apart. During this period the drop moves towards the centre of the channel, but slowly; here we quantify the time scale of this motion, verify that it is indeed much longer than that of the subsequent experiment, and discuss the maximum deformation of the channel in the doubly clamped state. 

Whilst it is possible to solve numerically our full problem with boundary conditions adjusted to account for both ends clamped (e.g.~by replacing \eqref{E:freeBC_dim} with $h(L, t) = H_0, h_x(L,t) = 0$, alongside the rest of~\eqref{E:beam_wet_dim}-\eqref{E:IC_dim}) it is more instructive to estimate the timescale by performing the analogue of the asymptotic calculation in the small $\hat{V}$ case presented in \S\ref{S:smallvhat}. (We have seen that in the full problem the time scale of motion is comparable to the results in the small $\hat{V}$ limit, and so expect the same to be the case when the boundary conditions are slightly modified.)

The analysis of \S\ref{S:smallvhat} shows that, when the end at $x = L$ is free, the leading meniscus evolves according to 
\begin{equation}\label{E:init_cond1}
\frac{\mathrm{d}\xright}{\mathrm{d}t} = \frac{\gamma^2 \cos^2 \theta \Delta XL^2}{3 \mu B H_0} f(\xright/L),
\end{equation}
where $f$ is the dimensionless function $f = f_f(s)= s^2 /2$. Similar analysis for the case where both ends are clamped results in a modified version of \eqref{E:init_cond1} with
\begin{equation}
f = f_c(s) = \frac{s^2 (2s-1)(s-1)^2}{2}.
\end{equation}
Note that $f_c$ is anti-symmetric about $s = 1/2$ and has $f_c(1/2) = 0$ so the system is in equilibrium when the drop sits at the centre of the channel, as expected; with the drop in this position, the beam displacement at its centre-point is maximised, and takes the value $1-h(1/2,t) \approx 0.005$. (Note that, regardless of drop position, the maximum beam displacement in the doubly clamped state is at $x= 1/2$).  This means the maximum beam displacement in the doubly clamped state is always less that 1$\%$ of the channel height; this justifies our use of an undeformed channel as initial condition  in \S~\ref{S:dim_math_model}.

\begin{figure}[h]
\centering
 \includegraphics[scale=0.5]{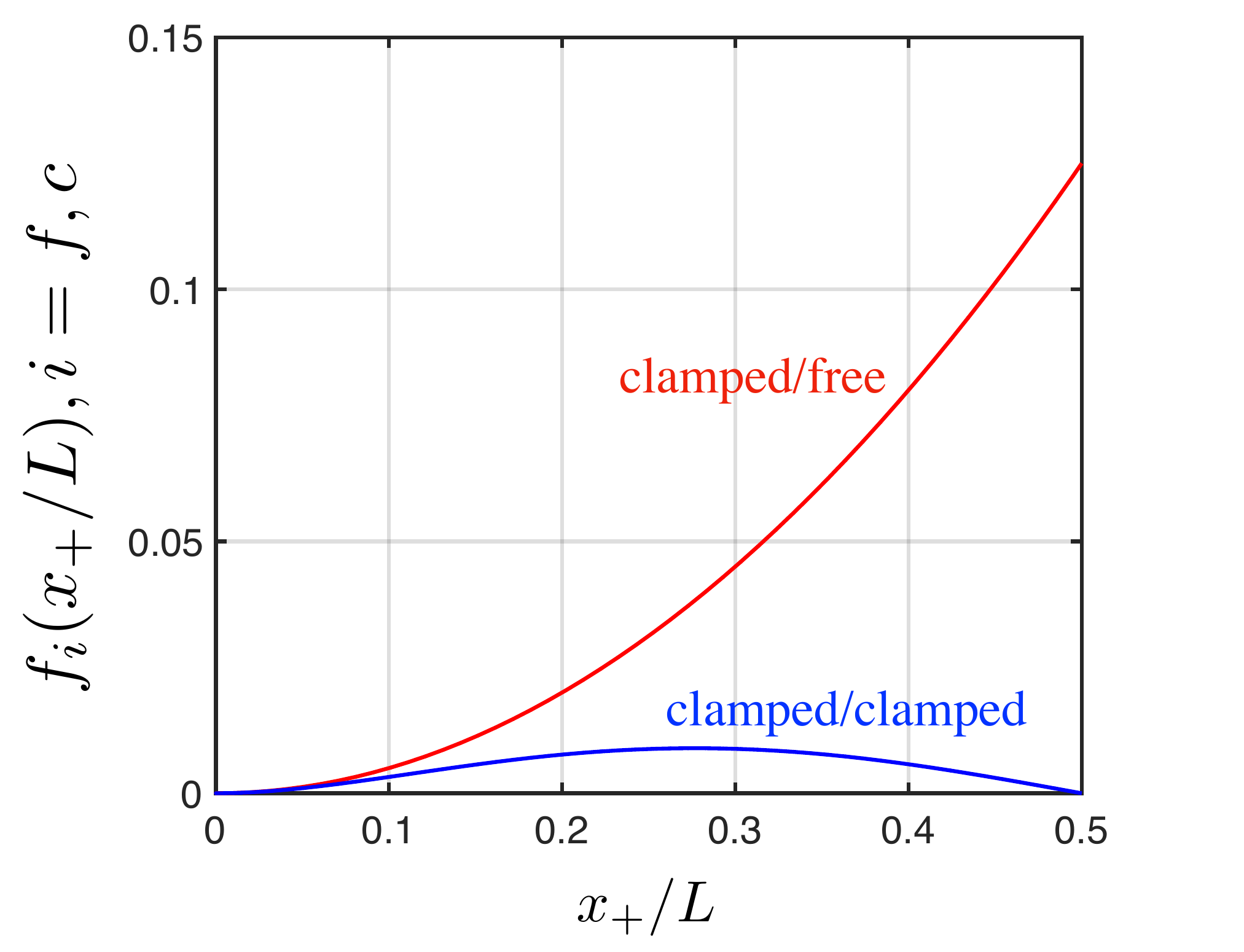}
\caption{\label{fig:init_cond} The dimensionless meniscus speeds with the far end free or clamped cases $f_i(x), \, i = f,c$, respectively, as defined in the main text.  The meniscus speed for the clamped far end (blue curve) is significantly slower than that in with the free far end (red curve).}
\end{figure}

Figure~\ref{fig:init_cond} compares the dimensionless functions $f_f$ and $f_c$; we see that $f_f(s)>f_c(s)$ throughout $|s|<1/2$, and conclude that motion in the small $\hat{V}$ case is significantly faster when the end at $x = L$ is free, rather than clamped. In particular, for $\xright/L > 0.37$, we predict the motion in the clamped case to be at least an order of magnitude slower than that with a free end. 

To estimate the time scale of the motion, we note that $f_c(s) \sim (s-1/2)/16$ as $s \to 1/2$. Therefore, with the far end clamped, the drop will approach the equilibrium at $\xright/L = 1/2$ like 
\begin{equation}
\frac{\xright}{L} - \frac{1}{2} \sim \exp\left(-\frac{t}{t_0}\right), \qquad t_0 = \frac{48 \mu B H_0}{\gamma^2 \cos^2 \theta \Delta X L}.
\end{equation}
Using typical experimental values, we find $t_0$ to be on the order of several minutes; this time scale is significantly longer than the time spent with the far end clamped ($\mathcal{O}(10\mathrm{s})$). (A similar estimate can be found by considering the time taken for the drop to move a length comparable to $L/2$ if it were to move at a constant speed corresponding to $\max_{0 < s<1/2} f_c(s) = \sqrt{5}/250$). We therefore conclude that the motion prior to release of the free end may be neglected.

%\bibliography{PRL_bib}